\documentclass[aps,showpacs,preprint,preprintnumbers,amsmath,amssymb,nofootinbib,contrib]{revtex4}
\usepackage{amssymb}
\usepackage{epsfig}
\usepackage{epsfig}
\usepackage{graphicx}% Include figure files
\usepackage{dcolumn}% Align table columns on decimal point
\usepackage{bm}% bold math
\def\beq{\begin{equation}}
\def\eeq{\end{equation}}

\newcommand{\bea}{\begin{eqnarray}}
\newcommand{\eea}{\end{eqnarray}}

\begin{document}

%\preprint{\parbox{1.2in}{\noindent arXiv:110 }}

\title{$\Omega_b$ semi-leptonic weak decays}

\author{Ming-Kai Du and Chun Liu}
\affiliation{
Institute of Theoretical Physics, Chinese Academy of Sciences,
and State Key Laboratory of Theoretical Physics, \\
P.O. Box 2735, Beijing 100190, China }
\email{mkdu@itp.ac.cn, liuc@mail.itp.ac.cn}

\begin{abstract}
$\Omega_b \rightarrow \Omega_c^{(*)}$ semi-leptonic decays are studied
in details.  Relevant helicity amplitudes are written down.  Both
unpolarized and polarized $\Omega_b$ cases are considered. Decay angular
distributions, asymmetry parameters and semileptonic decay rates are
calculated, with numerical results using leading order results of the
large $N_c$ heavy quark effective theory.
\end{abstract}



\maketitle

\section{Introduction}

Heavy baryons can be a good application ground of QCD.  They reveal some
important features of the heavy quark physics.  Data on heavy baryons
have been accumulating by experiments of LHC and Tevatron, as well as by
previously LEP, LEPII and B-factories.  Detailed theoretical analysis
are necessary.  The $\Lambda_b$ baryon has been studied considerably.
For an example, the $\Lambda_b \rightarrow \Lambda_c$ semileptonic decay
was analyzed thoroughly in Refs.
\cite{polarization effects, polarization effects2, lambda-b2, lambda-b3}
in terms of decay rates, distributions and various asymmetry parameters.

Although established for over 35 years, QCD's nonperturbative aspects
are still not fully understood, which render us from precise
calculations for the hadron physics.  For heavy hadrons containing a
single heavy quark, the heavy quark effective theory (HQET)
\cite{hqet1,hqet2} is the right QCD, which correctly factorizes the
perturbatively calculable part out from hadronic matrix elements of weak
currents in a simple and systematic way.  The really tough task lies in
calculating the nonperturbative part which is the universal Isgur-Wise
functions.  They can only be calculated by some nonperturbative methods
of QCD, like the large $N_c$ QCD \cite{large n}.

In this paper, $\Omega_b$ baryon semileptonic weak decays are studied.
The $\Omega_b$ baryon was discovered by Tevatron experiments
\cite{d0-cdf}, via its 2-body nonleptonic decay
$\Omega_b\to J/\Psi\Omega^-$.  In terms of the valence quark content, it
is made of $b-s-s$.  Unlike B-mesons or charm hadrons, b-baryons cannot
be produced at B-factories, they have been only produced at LEP,
Tevatron and LHC.  It would be a stable particle if the electroweak
interaction were shut down.  While the process
$\Omega_b\to J/\Psi\Omega^-$ is the most appropriate for determining the
$\Omega_b$ mass, the weak interaction properties of the $\Omega_b$
baryon cannot be precisely extracted out, because nonleptonic decays are
subjected to a large nonperturbative QCD uncertainty.  They are a lot
cleaner in the semileptonic decays $\Omega_b\to\Omega_c^{(*)}l\nu$ which
are not CKM suppressed.  In the near future, more data on $\Omega_b$
will be obtained by the Tevatron and LHCb experiments.  Furthermore, the
planning $Z$ factory \cite{z} can also produce a large amount of
$\Omega_b$ data.  In the $Z$ factory, $Z$ is polarized, $\Omega_b$
coming out from $Z$ is also polarized.  All these make it viable to
analyze the $\Omega_b$ semileptonic decays experimentally.
Theoretically semileptonic decays are simply parameterized in terms of
form factors which contain all the nonperturbative QCD effects.
With the help of the HQET, there are only two universal Isgur-Wise
functions at the leading order of heavy quark expansion in the
$\Omega_b \rightarrow \Omega_c^{(*)}$ transitions \cite{form factors}.
These Isgur-Wise functions can be further calculated in the large $N_c$
QCD \cite{I-W, large n2}.  This is partly based on the observation of
the light-quark spin-flavor symmetry in the large $N_c$ limit
\cite{large n3}.

We will perform a detailed analysis considering polarization effects of
the decays.  Our analysis follows the way of K\"orner and Kr\"amer
\cite{polarization effects} who analyzed $\Lambda_b$ semileptonic decays.
The technique of helicity amplitudes is adopted which can be found in
\cite{helicity1,helicity2}.  For obtaining detailed information of the
$\Omega_b$ decays, all kinds of observables are calculated, although
some of them are not practically measurable in the current stage.
Nevertheless in such a systematic way, the semileptonic decay branching
ratio and spectrum are also obtained at last.  In Sect. II, helicity
amplitudes are written down for analyzing the
$\Omega_b \rightarrow \Omega_c^{(*)}$ weak decays.  Decay distributions
and various asymmetry parameters are calculated in Sect. III.  The decay
rates are presented in Sect. IV. In Sect. V, we summarize the results.

\section{Form factors and Helicity amplitudes}
%%%%%%%%%%%%%%%%%%%%%%%%%%%%%%%%%%%%%%%%%%%%%%%%%%%%5%%%5
\subsection{Form factors}
%%%%%%%%%%%%%%%%%%%%%%%%%%%%%%%%%%%%%%%%%%%%%%%%%%

The hadronic matrix elements of the weak currents
$V_\mu\equiv \bar{c}\gamma_\mu b$ and
$A_\mu\equiv \bar{c}\gamma_\mu\gamma_5 b$ can be parametrized by
fourteen form factors which are defined as below \cite {expressions},
\bea
\langle\Omega_c(v',s')|V^\mu|\Omega_b(v)\rangle&=&
\bar{u}(v',s')(F_1\gamma^{\mu}+F_2v^{\mu}+F_3v'^{\mu})u(v,s);
\nonumber\\
\langle\Omega_c(v',s')|A^\mu|\Omega_b(v)\rangle&=&
\bar{u}(v',s')(G_1\gamma^{\mu}+G_2v^{\mu}+G_3v'^{\mu})\gamma^5 u(v,s);
\nonumber \\
\langle\Omega^*_c(v',s')|V^\mu|\Omega_b(v)\rangle&=&
\bar{u}_{\lambda}(v',s')(N_1v^{\lambda}\gamma^{\mu}
+N_2v^{\lambda}v^{\mu}+N_3v^{\lambda}v'^{\mu}
+N_4g^{\lambda\mu})\gamma^5u(v,s);\nonumber \\
\langle\Omega^*_c(v',s')|A^\mu|\Omega_b(v)\rangle&=&
\bar{u}_{\lambda}(v',s')(K_1v^{\lambda}\gamma^{\mu}
+K_2v^{\lambda}v^{\mu}+K_3v^{\lambda}v'^{\mu}+K_4g^{\lambda\mu})u(v,s).
\label {eq:general}
\eea
where $u_{\lambda}$ is the Rarita-Schwinger spinor for the $\Omega^*_c$.
It is convenient to redefine some of the form factors as below:
\bea
F'_2=\frac 1 2 \Big ( \frac {F_2} {M_1}+\frac {F_3} {M_2}\Big ),\quad
F'_3=\frac 1 2 \Big ( \frac {F_2} {M_1}-\frac {F_3} {M_2}\Big );
\nonumber\\[0.5cm]
G'_2=\frac 1 2 \Big ( \frac {G_2} {M_1}+\frac {G_3} {M_2}\Big ),\quad
G'_3=\frac 1 2 \Big ( \frac {G_2} {M_1}-\frac {G_3} {M_2}\Big );
\nonumber\\[0.5cm]
N'_2=\frac 1 2 \Big ( \frac {N_2} {M_1}+\frac {N_3} {M'_2}\Big ),\quad
N'_3=\frac 1 2 \Big ( \frac {N_2} {M_1}-\frac {N_3} {M'_2}\Big );
\nonumber\\[0.5cm]
K'_2=\frac 1 2 \Big ( \frac {K_2} {M_1}+\frac {K_3} {M'_2}\Big ),\quad
K'_3=\frac 1 2 \Big ( \frac {K_2} {M_1}-\frac {K_3} {M'_2}\Big ),
\eea
where $M_1$ is the $\Omega_b$ mass, $M_2$ and $M'_2$ masses of
$\Omega_c$ and $\Omega^*_c$ masses, respectively, while $M_1=6.071$ GeV,$
M_2=2.695$ GeV, and $M'_2=2.770$ GeV \cite{mass}.  For simplicity, we
shall neglect lepton masses.  In this case,
$F'_3, ~G'_3, ~F'_3, ~N'_3$ and $K'_3$ have no contribution to the
decays.

In the HQET, according to the standard tensor method \cite{form factors},
we denote the $\Omega_Q^{(*)}$ states by $\Omega_Q^M$, where $M=1$ is
for $\Omega_Q$ and $M=2$ for $\Omega_Q^*$.  Then the tensor fields
describing the $\Omega_Q^M$ states are $B_\mu^M$,
\bea
B_\mu^1(v,s)=\frac{1}{\sqrt{3}}(\gamma_\mu+v_\mu)\gamma^5u(v,s), \qquad
B_\mu^2(v,s)=u_\mu(v,s)\,.
\eea

To the leading order of heavy quark expansion, the fourteen form factors
are reduced into two Isgur-Wise functions \cite{form factors},
\bea
\langle\Omega_c^M|\bar{h}^{(c)}\Gamma h^{(b)}|\Omega_b^N \rangle&=&C
\bar{B}_\mu^M\Gamma B^N [-g^{\mu\nu}\xi_1(\omega)
+v^\mu v'^\nu \xi_2(\omega)] \, ,
\eea
\bea
C=\bigg[\frac{\alpha_s(m_b)}{\alpha_s(m_c)}\bigg]^{-6/25}=1.1 \, .
\eea
where $\omega=v\cdot v'$, and $C$ is the QCD perturbative leading logarithm
correction, which has been evaluated at the scale $\mu=m_c$. The fourteen
form factors are then expressed
as below \cite{expressions},
\begin{eqnarray}
\begin{array}{rrrrrrrrrrr}
F_1&=& \displaystyle
\frac{-\omega}{3}\xi_1&+&\displaystyle\frac{\omega^2-1}{3}\xi_2,&\qquad&
G_1&=&\displaystyle \frac{-\omega}{3}\xi_1&
+&\displaystyle\frac{\omega^2-1}{3}\xi_2\\[0.5 cm]
F_2&=& \displaystyle
\frac{2}{3}\xi_1&+&\displaystyle \frac{2(1-\omega)}{3}\xi_2,&\qquad&
G_2&=&\displaystyle \frac{2}{3}\xi_1&
+&\displaystyle \frac{-2(1+\omega)}{3}\xi_2\\[0.5 cm]
F_3&=& \displaystyle
\frac{2}{3}\xi_1&+&\displaystyle \frac{2(1-\omega)}{3}\xi_2,&\qquad&
G_3&=&\displaystyle \frac{-2}{3}\xi_1&
+&\displaystyle \frac{2(1+\omega)}{3}\xi_2\\[0.5 cm]
N_1&=& \displaystyle \frac{-1}{\sqrt{3}}\xi_1&
+&\displaystyle \frac{\omega-1}{\sqrt3}\xi_2,&\qquad&
K_1&=&\displaystyle \frac{-1}{\sqrt{3}}\xi_1&
+&\displaystyle \frac{\omega+1}{\sqrt3}\xi_2\\[0.5 cm]
N_2&=& 0&\quad&,&\qquad&K_2&=&0&\quad&\\[0.5 cm]
N_3&=& \displaystyle
0&+&\displaystyle \frac{2}{\sqrt{3}}\xi_2,&\qquad&K_3&=&0&+&
\displaystyle \frac{-2}{\sqrt{3}}\xi_2\\[0.5 cm]
N_4&=& \displaystyle
\frac{-2}{\sqrt{3}}\xi_1&+&0,&\qquad&K_4&=&
\displaystyle \frac{2}{\sqrt{3}}\xi_1&+&0 \, ,
\end{array}
\end{eqnarray}
it is at this stage that nonperturbation methods are needed.  In the
large $N_c$ limit, these two Isgur-Wise functions are related to that of
$\langle\Lambda_c|\bar{h}^{(c)}\Gamma h^{(b)}|\Lambda_b \rangle$.
While
$\langle\Lambda_c|\bar{h}^{(c)}\Gamma h^{(b)}|\Lambda_b \rangle=\eta\bar{u}_c\Gamma u_b$,
the relations are \cite{relation,large n2}:
\bea
\eta(\omega)=\xi_1(\omega)=(\omega +1)\xi_2(\omega)\,.
\eea
Furthermore, in the large $N_c$ limit, $\eta$ is predicted as \cite{I-W}:
\bea
\eta(\omega)&=&0.99\exp[-1.3(\omega-1)] \,.
\eea

%%%%%%%%%%%%%%%%%%%%%%%%%%%%%%%%%%%%%%%%%%%%%%%%%%%%%%%%%%%%%%%%5
\subsection{Helicity amplitudes}

Following the way of Ref. \cite{polarization effects} for
$\Lambda_b\rightarrow\Lambda_c l\bar{\nu}$ decays, we analyze
$\Omega_b\rightarrow\Omega_c^{(*)}l\bar{\nu}$ semileptonic
decays.  It is convenient to regard the decay as two-successive decays
$\Omega_1\rightarrow\Omega_2+W_{\rm off-shell}$ and
$W_{\rm off-shell}\rightarrow\ell+\bar{\nu}$.  We denote helicity amplitudes
of $\Omega_b\rightarrow\Omega_c+\ell+\bar{\nu}$ as
$H_{\lambda_2\lambda_W}^{V,A}$, and that of
$\Omega_b\rightarrow\Omega_c^* +\ell+\bar{\nu}$ as
$H_{\lambda_2\lambda_W}^{'V,A}$, where $\lambda_2$ and $\lambda_W$ are
helicities of the daughter baryon and the off-shell $W$-boson.  These
amplitudes can be expressed by our redefined form factors as:
\bea
\begin{array}{lllllll}
\sqrt{q^2}H_{1/2\ 0}^V&=&\sqrt{Q_-}[(M_1+M_2)F_1+F_2'Q_+]&,&
H_{1/2\ 1}^V&=&-\sqrt{2Q_-}F_1;\\[0.5 cm]
\sqrt{q^2}H_{1/2\ 0}^A&=&\sqrt{Q_+}[(M_1-M_2)G_1-G_2'Q_-]&,&
H_{1/2\ 1}^A&=&-\sqrt{2Q_+}G_1;\\[1.0 cm]
\end{array}
\eea
and
\bea
\begin{array}{ccl}
\sqrt{q'^2}H_{1/2\ 0}^{'V}&=& \displaystyle
\sqrt{\frac{2}{3}}\frac{p'}{M'_2}\sqrt{Q'_+}
[(M_1-M'_2)N_1-N_2'Q_-]\\[0.5 cm]
&&\displaystyle-\sqrt{\frac23}\sqrt{Q'_-}\bigg[\frac {Q_-}{2M'_2}+(M_1-M'_2)\bigg]N_4
\\[0.5 cm]
\sqrt{q'^2}H_{1/2\ 0}^{'A}&=& \displaystyle
\sqrt{\frac{2}{3}}\frac{p'}{M'_2}\sqrt{Q'_-}
[(M_1+M'_2)K_1+K_2'Q_+]\\[0.5 cm]
&&\displaystyle+\sqrt{\frac23}\sqrt{Q'_+}\bigg[\frac {Q_+}{2M'_2}-(M_1+M'_2)\bigg]K_4
\\[0.5 cm]
H_{1/2\ 1}^{'V}&=& \displaystyle
\sqrt{\frac13}\sqrt{Q'_-}\bigg[N_4-N_1\frac{Q'_+}{M_1M'_2}\bigg]\\[0.5 cm]
H_{1/2\ 1}^{'A}&=& \displaystyle
\sqrt{\frac13}\sqrt{Q'_+}\bigg[K_4-K_1\frac{Q'_-}{M_1M'_2}\bigg]\\[0.5 cm]
H_{3/2\ 1}^{'V}&=&-N_4\sqrt{Q'_-}\\[0.5 cm]
H_{3/2\ 1}^{'A}&=&K_4\sqrt{Q'_+}\, ,\\[0.5 cm]
\end{array}
\eea
where $Q^{(\prime)}_\pm=(M_1\pm M^{(\prime)}_2)^2-q^{(\prime)2}$ and
$q^{(\prime)^\mu}(W)=(q^{(\prime)0},0,0,-p^{(\prime)})$ while
$p^{(\prime)}=\sqrt{Q^{(\prime)}_+Q^{(\prime)}_-}/2M_1$ and
$q^{(\prime)0}=(M_1^2-M^{(\prime)2}_2+q^{(\prime)2})/2M_1$.  Other
helicity amplitudes can be obtained via using the parity relations:
\bea
H_{-\lambda_2-\lambda_W}^{V(A)}=+(-)H_{\lambda_2\lambda_W}^{V(A)}\, .
\eea

%%%%%%%%%%%%%%%%%%%%%%%%%%%%%%%%%%%%%%%%%%%%%%%%%%%%%%%%%%%%%
\section{Angular distributions and asymmetry parameters}
%%%%%%%%%%%%%%%%%%%%%%%%%%%%%%%%%%%%%%%%%%%%%%%%%%%%%%%%%%%%

Unpolarized and polarized $\Omega_b$ decays will be considered,
respectively.  And in case of the $\Omega_b\rightarrow \Omega_c$
transition, the cascade nonleptonic weak decay $\Omega_c\rightarrow a+b$
(for example $\Omega_c\rightarrow \Omega+\pi$ \cite{mass}) will be taken
into account, where $a$ has spin 1/2, and $b$ is a spin zero particle.
While in the case of $\Omega_b\rightarrow \Omega^*_c$ transition, we
will not further consider $\Omega_c^*$ cascade decays which are either
strong or radiative decays \cite{mass} and therefore will not produce
the asymmetry factors.

%%%%%%%%%%%%%%%%%%%%%%%%%%%%%%%%%%%%%%%%%%%%%%%%%%%%%%%%%%%%
\subsection{Unpolarized $\Omega_b$ decay}
%%%%%%%%%%%%%%%%%%%%%%%%%%%%%%%%%%%%%%%%%%%%%%%%%%%%

For that $\Omega_b$ is unpolarized, it is convenient to introduce the
correlation density matrix first, which is given by
\bea
\rho_{\lambda_2\lambda_W;\lambda'_2\lambda'_W}&=&
H_{\lambda_2\lambda_W}H^*_{\lambda'_2\lambda'_W}\,.
\eea
With this density matrix, using the methods of Refs.
\cite{helicity1, helicity2, helicity3} and ignoring lepton masses, we
obtain the angular distribution for the whole decay
$\Omega_b\rightarrow\Omega_c (\rightarrow a+b)+W(\rightarrow\ell+\bar{\nu})$:
\bea \label {eq :distribution}
\frac{\mathrm{d}\Gamma}{\mathrm{d}\omega\mathrm{d}\cos\Theta
\mathrm{d}\chi\mathrm{d}\cos\Theta_\Omega}&=&
Br(\Omega_c\rightarrow a+b)\frac{G^2}{(2\pi)^4}|V_{cb}|^2
q^2\sqrt{\omega^2-1}\frac{M^2_2}{24M_1}\nonumber \\[0.5cm]
& &\times\bigg(\frac{3}{8}(1+\cos\Theta)^2\big|H_{1/2\ 1}\big|^2
(1+\alpha_\Omega\cos\Theta_\Omega)\nonumber\\[0.5cm]
& &+\frac{3}{8}(1-\cos\Theta)^2\big|H_{-1/2\ -1}\big|^2
(1-\alpha_\Omega\cos\Theta_\Omega)\nonumber\\[0.5cm]
& &+\frac{3}{4}\sin\Theta^2\big|H_{1/2\ 0}\big|^2
(1+\alpha_\Omega\cos\Theta_\Omega)\nonumber\\[0.5cm]
& &+\frac{3}{4}\sin\Theta^2\big|H_{-1/2\ 0}\big|^2
(1-\alpha_\Omega\cos\Theta_\Omega)\nonumber\\[0.5cm]
& &-\frac{3}{2\sqrt{2}}\alpha_\Omega\cos\chi\sin\Theta
\sin\Theta_\Omega[(1+\cos\Theta)
\mathrm{Re}(H_{-1/2\ 0}H_{1/2\ 1}^*)]\nonumber\\[0.5cm]
& &-\frac{3}{2\sqrt{2}}\alpha_\Omega\cos\chi\sin\Theta
\sin\Theta_\Omega[(1-\cos\Theta)
\mathrm{Re}(H_{1/2\ 0}H_{-1/2\ -1}^*)]\bigg)\, ,\nonumber
\\
\eea
where the polar angle $\Theta$ is for $l$, $\Theta_\Omega$ for $a$, and
$\chi$ is the azimuthal angle.  These angles are illustrated in Fig.1
and Fig.2.  $G$ is the Fermi coupling and $V_{cb}$ is the
Cabibbo-Kobayashi-Maskawa (CKM) mixing matrix element.  The daughter
baryon $\Omega_c$ decays into $a$ and $b$ with a branching ratio
$Br(\Omega_c\rightarrow a+b)$ and decay asymmetry parameter
$\alpha_\Omega$.  $p$ is the momentum of $\Omega_c$ in the rest
reference frame of $\Omega_b$.  According to the results of
\cite{cp violation}, in Eq.(\ref {eq :distribution}) we have assumed all
the helicity amplitudes are real, since otherwise we will have to
include the effects of $CP$-violation.

Various angular distribution and asymmetry parameters of $\Omega_b$
semileptonic decays can now be obtained.  First, from
Eq.(\ref {eq :distribution}), by integrating other angles,
the polar angle distribution of the successive decay
$\Omega_c\rightarrow a+b$ is
\bea
\frac{\mathrm{d}\Gamma}
{\mathrm{d}\omega\mathrm{d}\cos\Theta_\Omega}\propto
1+\alpha_1\alpha_\Omega\cos\Theta_\Omega \, ,
\eea
where the asymmetry parameter $\alpha_1$ is defined as
\bea
\alpha_1=\frac{\big|H_{1/2\ 1}\big|^2-\big|H_{-1/2\ -1}\big|^2
+\big|H_{1/2\ 0}\big|^2-\big|H_{-1/2\ 0}\big|^2}
{\big|H_{1/2\ 1}\big|^2+\big|H_{-1/2\ -1}\big|^2+\big|H_{1/2\ 0}\big|^2
+\big|H_{-1/2\ 0}\big|^2} \,,
\eea
and the polar angle distribution of the decay
$W\rightarrow \ell+\bar{\nu}$ is
\bea
\frac{\mathrm{d}\Gamma}{\mathrm{d}\omega\mathrm{d}\cos\Theta}\propto
1+2\alpha _2\cos\Theta+\alpha_3\cos^2\Theta \, ,
\eea
where the parameters $\alpha_2$ and $\alpha_3$ are
\bea
\alpha_2&=&\frac{\big|H_{1/2\ 1}\big|^2-\big|H_{-1/2\ -1}\big|^2}
{\big|H_{1/2\ 1}\big|^2+\big|H_{-1/2\ -1}\big|^2
+2(\big|H_{1/2\ 0}\big|^2+\big|H_{-1/2\ 0}\big|^2)} \,,
\eea
\bea
\alpha_3=\frac{\big|H_{1/2\ 1}\big|^2+\big|H_{-1/2\ -1}\big|^2
-2(\big|H_{1/2\ 0}\big|^2+\big|H_{-1/2\ 0}\big|^2)}
{\big|H_{1/2\ 1}\big|^2+\big|H_{-1/2\ -1}\big|^2
+2(\big|H_{1/2\ 0}\big|^2+\big|H_{-1/2\ 0}\big|^2)} \,,
\eea
and the $\chi$ distribution is
\bea
\frac{\mathrm{d}\Gamma}{\mathrm{d}\omega\mathrm{d}\chi}
\propto 1-\frac{3\pi ^2}{32\sqrt{2}}\gamma\alpha_\Omega\cos\chi \, ,
\eea
where
\bea
\gamma=\frac{2\mathrm{Re}\Big(H_{-1/2\ 0}H_{1/2\ 1}^*
+H_{1/2\ 0}H_{-1/2\ -1}^*\Big)}
{\big|H_{1/2\ 1}\big|^2+\big|H_{-1/2\ -1}\big|^2
+\big|H_{1/2\ 0}\big|^2+\big|H_{-1/2\ 0}\big|^2} \,.
\eea

Up to now, all of the analysis in this section are model independent.
With the help of the large $N_c$ Isgur-Wise function given in Sec.II, we
can calculate all these asymmetry parameters numerically, the results are
listed in Table I.

\begin{figure}
\scalebox{0.5}[0.5]{\includegraphics{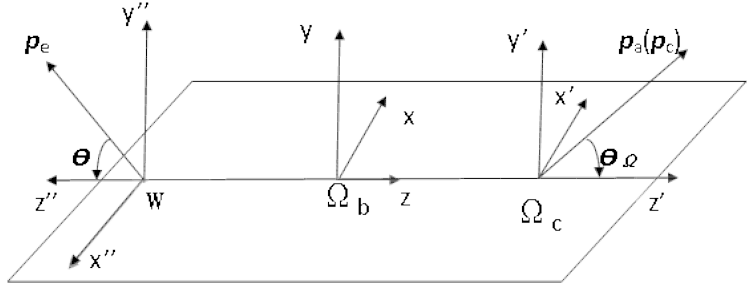}}
\caption {Definition of polar angles $\Theta_\Omega$ and $\Theta$,
both angles are defined in rest frames of the decaying particles.}
\end{figure}

\begin{figure}
\scalebox{0.4}[0.4]{\includegraphics{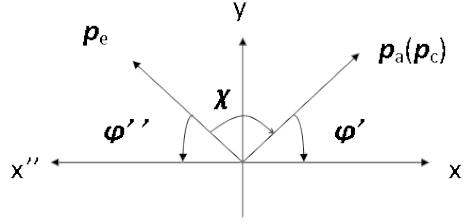}}
\caption {Definition of the azimuthal angle $\chi$ which is the one
between two cascade decay planes.}
\end{figure}

Next, let us turn to the analysis of the decay
$\Omega_b\rightarrow\Omega^*_c+W(\rightarrow \ell+\bar{\nu})$,
the procedure is analogous to the analysis of
$\Omega_b\rightarrow\Omega_c(\rightarrow a+b)+W(\rightarrow \ell+\bar{\nu})$,
we can get the angular distribution as the following:
\bea
\frac{\mathrm{d}\Gamma'}{\mathrm{d}\omega\mathrm{d}\cos\Theta}&=&
\frac{G^2}{(2\pi)^3}|V_{cb}|^2 q'^2\sqrt{\omega^2-1}\frac{M'^2_2}{12M_1}\nonumber \\[0.5cm]
& &\times\bigg(\frac{3}{8}(1+\cos\Theta)^2\big|H'_{3/2\ 1}\big|^2
+\frac{3}{8}(1-\cos\Theta)^2\big|H'_{-3/2\ -1}\big|^2\nonumber\\[0.5cm]
& &+\frac{3}{8}(1+\cos\Theta)^2\big|H'_{1/2\ 1}\big|^2
+\frac{3}{8}(1-\cos\Theta)^2\big|H'_{-1/2\ -1}\big|^2\nonumber\\[0.5cm]
& &+\frac{3}{4}\sin^2\Theta\big|H'_{1/2\ 0}\big|^2
+\frac{3}{4}\sin^2\Theta\big|H'_{-1/2\ 0}\big|^2 \bigg),
\eea
where the angle $\Theta$ has the same meaning as before. Again we can
get some asymmetry parameters. The polar angular distribution of the cascade decay of
$W\rightarrow \ell+\bar{\nu}$ is
\bea
\frac{\mathrm{d}\Gamma'}{\mathrm{d}\omega\mathrm{d}\cos\Theta}\propto
1+2\alpha'_1\cos\Theta+\alpha'_2\cos^2\Theta \, ,
\eea
where
\bea
\alpha'_1=\frac{\big|H'_{3/2\ 1}\big|^2-\big|H'_{-3/2\ -1}\big|^2
+\big|H'_{1/2\ 1}\big|^2-\big|H'_{-1/2\ -1}\big|^2}
{\big|H'_{3/2\ 1}\big|^2+\big|H'_{-3/2\ -1}\big|^2
+\big|H'_{1/2\ 1}\big|^2+
\big|H'_{-1/2\ -1}\big|^2+2\big(\big|H'_{1/2\ 0}\big|^2
+\big|H'_{-1/2\ 0}\big|^2\big)} \,,
\eea
\bea
\alpha'_2=\frac{\big|H'_{3/2\ 1}\big|^2+\big|H'_{-3/2\ -1}\big|^2
+\big|H'_{1/2\ 1}\big|^2+
\big|H'_{-1/2\ -1}\big|^2-2\big(\big|H'_{1/2\ 0}\big|^2
+\big|H'_{-1/2\ 0}\big|^2\big)}
{\big|H'_{3/2\ 1}\big|^2+\big|H'_{-3/2\ -1}\big|^2
+\big|H'_{1/2\ 1}\big|^2+
\big|H'_{-1/2\ -1}\big|^2+2\big(\big|H'_{1/2\ 0}\big|^2
+\big|H'_{-1/2\ 0}\big|^2\big)} \,.
\eea\\

All the numerical results of these asymmetry parameters are listed in
Table I.

%%%%%%%%%%%%%%%%%%%%%%%%%%%%%%%%%%%%%%%%%%%%%%%%%%%
\subsection{Polarized $\Omega_b$ decay}
%%%%%%%%%%%%%%%%%%

In this sub-section, the decays of a polarized $\Omega_b$ will be
analyzed, since in the proposed Z-factory \cite{z},  the produced
bottom quarks will be polarized.
It is reasonable to assume the $\Omega_b$ will also be polarized in Z-factory.
Two new decay angles will be introduced, $\Theta_P$ and
$\chi_p$, where $P$ denotes the polarization vector of the parent baryon
$\Omega_b$, the angles involved are shown in Fig.3 and Fig.4.

For the decay
$\Omega_b\rightarrow \Omega_c (\rightarrow a+b)+W(\rightarrow \ell+\bar{\nu})$,
the density matrix is now the following,
\bea
\rho_{1/2\ 1/2}&=&\big|H_{1/2\ 1}\big|^2(1-P\cos{\Theta_P})
+\big|H_{1/2\ 0}\big|^2(1+P\cos{\Theta_P}) \,, \nonumber\\
\rho_{1/2\ -1/2}&=&\rho_{-1/2\ 1/2}=P\sin{\Theta_P}
\mathrm{Re}(H_{1/2\ 0}H_{-1/2\ 0}^*) \,, \nonumber\\
\rho_{-1/2\ -1/2}&=&\big|H_{-1/2\ -1}\big|^2(1+P\cos{\Theta_P})
+\big|H_{-1/2\ 0}\big|^2(1-P\cos{\Theta_P}) \,.
\eea

\begin{figure}
\scalebox{0.5}[0.5]{\includegraphics{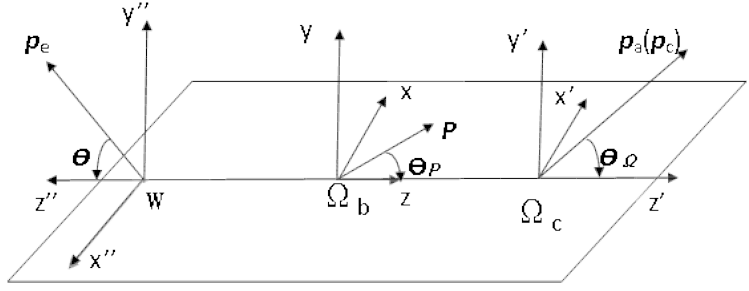}}
\caption {Definition of polar angles $\Theta_\Omega$ and $\Theta_P$,
where the polarization vector $P$ is in the y-z plane. }
\end{figure}

\begin{figure}
\scalebox{0.4}[0.4]{\includegraphics{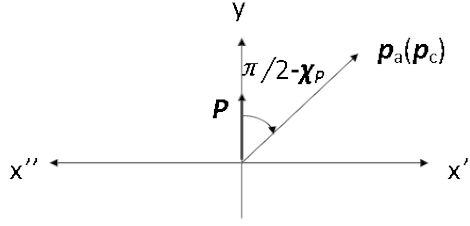}}
\caption {Definition of azimuthal angle $\chi_P$.}
\end{figure}
After integrating the angles of the leptons out, the whole angle
distribution is obtained:
\bea
&&\frac{\mathrm{d}\Gamma}{\mathrm{d}\omega\mathrm{d}\cos\Theta_P
\mathrm{d}\chi_P\mathrm{d}\cos\Theta_\Omega}
=Br(\Omega_c\rightarrow a+b)\frac{G^2}{(2\pi)^4}|V_{cb}|^2
q^2\sqrt{\omega^2-1}\frac{M^2_2}{48M_1}\nonumber \\[0.5cm]
&&\times\bigg(\big|H_{1/2\ 1}\big|^2+\big|H_{-1/2\ -1}\big|^2
+\big|H_{1/2\ 0}\big|^2+
\big|H_{-1/2\ 0}\big|^2\nonumber\\[0.5cm]
&&+\alpha_\Omega\cos{\Theta_\Omega}(\big|H_{1/2\ 1}\big|^2
-\big|H_{-1/2\ -1}\big|^2+
\big|H_{1/2\ 0}\big|^2-\big|H_{-1/2\ 0}\big|^2)x\nonumber\\[0.5cm]
&&+P\alpha_\Omega\cos{\Theta_P}(-\big|H_{1/2\ 1}\big|^2
+\big|H_{-1/2\ -1}\big|^2+
\big|H_{1/2\ 0}\big|^2-\big|H_{-1/2\ 0}\big|^2)\nonumber\\[0.5cm]
&&+P\alpha_\Omega\cos{\Theta_\Omega}\cos{\Theta_P}
(-\big|H_{1/2\ 1}\big|^2-
\big|H_{-1/2\ -1}\big|^2+\big|H_{1/2\ 0}\big|^2
+\big|H_{-1/2\ 0}\big|^2)\nonumber\\[0.5cm]
&&+P\alpha_\Omega\sin{\Theta_\Omega}\sin{\Theta_P}\cos{\chi_P}2
\mathrm{Re}(H_{1/2\ 0}H^*_{-1/2\ 0})\bigg) \,.
\eea
Then the $\Theta_P$ angle distribution is
\bea
\frac{\mathrm{d}\Gamma}{\mathrm{d}\omega\mathrm{d}\cos\Theta_P}\propto
1-\alpha _PP\cos\Theta_P \, ,
\eea
where
\bea
\alpha_P=\frac{\big|H_{1/2\ 1}\big|^2-\big|H_{-1/2\ -1}\big|^2
-\big|H_{1/2\ 0}\big|^2+\big|H_{-1/2\ 0}\big|^2}
{\big|H_{1/2\ 1}\big|^2+\big|H_{-1/2\ -1}\big|^2+\big|H_{1/2\ 0}\big|^2
+\big|H_{-1/2\ 0}\big|^2} \,.
\eea
And the $\chi_P$ distribution is
\bea
\frac{\mathrm{d}\Gamma}{\mathrm{d}\omega\mathrm{d}\chi}\propto 1
-\frac{\pi ^2}{16}P\gamma_P\alpha_P\cos\chi \, ,
\eea
where
\bea
\gamma_P=\frac{2\mathrm{Re}\Big(H_{1/2\ 0}H_{-1/2\ 0}^*\Big)}
{\big|H_{1/2\ 1}\big|^2+\big|H_{-1/2\ -1}\big|^2
+\big|H_{1/2\ 0}\big|^2+\big|H_{-1/2\ 0}\big|^2} \, .
\eea
The numerical results of these asymmetry parameters are shown in Table I.

For the decay
$\Omega_b\rightarrow \Omega^*_c+W(\rightarrow \ell+\bar{\nu})$,
after integrating the lepton angles out, there are no such
two asymmetry factors.

\begin{table}[h]
\caption{Asymmetry parameters}
\begin{tabular}{|l|cccccccc|}
\hline\hline
&$\alpha_1$ &$\alpha_2$ &$\alpha_3$ &$\alpha_P$ &$\gamma$ &$\gamma_P$ &$\alpha'_1$&$\alpha'_2$ \\
\hline
$\omega$ =1
&0         &0          &0          &0          &0.943   &-1/3 &0          &0 \\
mean-value
&0.522      &-0.04     &-0.751     &-0.626     &0.478   &0.468   &-0.132   &-0.363 \\
\hline\hline
\end{tabular}
\end{table}

%%%%%%%%%%%%%%%%%%%%%%%%%%%%%%%%%%%%%%%%%
\section{The decay rates}
%%%%%%%%%%%%%%%%%%%%%%%%%%%%%%%%%%%%%%%%
To be more concrete, we can now calculate the differential decay rates.
Neglecting the lepton mass, the $\Omega_b\rightarrow\Omega_c\ l\ \bar\nu$
differential decay rate can be expressed in terms of the helicity
amplitudes as

\bea
\frac{\mathrm{d}\Gamma(\omega)}{\mathrm{d}\omega}=
&& \frac{G^2}{(2\pi)^3}|V_{cb}|^2
q^2\sqrt{\omega^2-1}\frac{M^2_2}{12M_1}C^2\nonumber \\
&&\times\Big[\big|H_{1/2\ 1}\big|^2+\big|H_{-1/2\ -1}\big|^2
+\big|H_{1/2\ 0}\big|^2+\big|H_{-1/2\ 0}\big|^2\Big] \, ,
\eea
where $q^2=M^2_1+M^2_2-2M_1M_2\omega$.

For the decay of $\Omega_b\rightarrow\Omega^*_c\ l\ \bar\nu$, we
have :
\bea
\frac{\mathrm{d}\Gamma'(\omega)}{\mathrm{d}\omega}= &&
\frac{G^2}{(2\pi)^3}|V_{cb}|^2 q'^2\sqrt{\omega^2-1}
\frac{M'^2_2}{12M_1}C^2\nonumber \\
&&\times\Big[\big|H'_{3/2\ 1}\big|^2+\big|H'_{-3/2\ -1}\big|^2
+\big|H'_{1/2\ 1}\big|^2+\big|H'_{-1/2\ -1}\big|^2
+\big|H'_{1/2\ 0}\big|^2+\big|H'_{-1/2\ 0}\big|^2\Big] \, ,\nonumber \\
\eea

where $q'^2=M^2_1+M^2_3-2M_1M_3\omega$, and the above distributions
are plotted in Fig. 5 and Fig. 6. All the results are consistent
with \cite {decay rate1,decay rate2, decay rate3} when expressed in terms of
form factors.

By inputting the form factors discussed
in Sect. II, numerical results can be obtained. We have taken
$G=1.16637\times10^{-5}\,\mathrm{GeV^{-2}}$
and $|V_{cb}|=40.6\times 10^{-3}$ \cite{mass}. The results are:
\bea
&\Gamma(\Omega_b\rightarrow\Omega_c\ l\ \bar\nu)=1.686\times10^{-14} \mathrm{GeV} ,&
\nonumber \\
&{\rm B}(\Omega_b\rightarrow\Omega_c\ l\ \bar\nu)=2.82\%.&
\eea

\bea
&\Gamma(\Omega_b\rightarrow\Omega^*_c\ l\ \bar\nu)=3.482\times10^{-14} \mathrm{GeV} ,&
\nonumber \\
&{\rm B}(\Omega_b\rightarrow\Omega^*_c\ l\ \bar\nu)=5.82\%.&
\eea

The second width is about twice as large as the first one,
this can be understood easily when we consider the Clebsch-Gordan coefficients.
Note that we have obtained the above results by taking two
approximations: heavy quark limit and large $N_c$ limit.
In the near future, these results can
be tested at the LHCb experiment.
\begin{figure}
\scalebox{0.8}[0.8]{\includegraphics{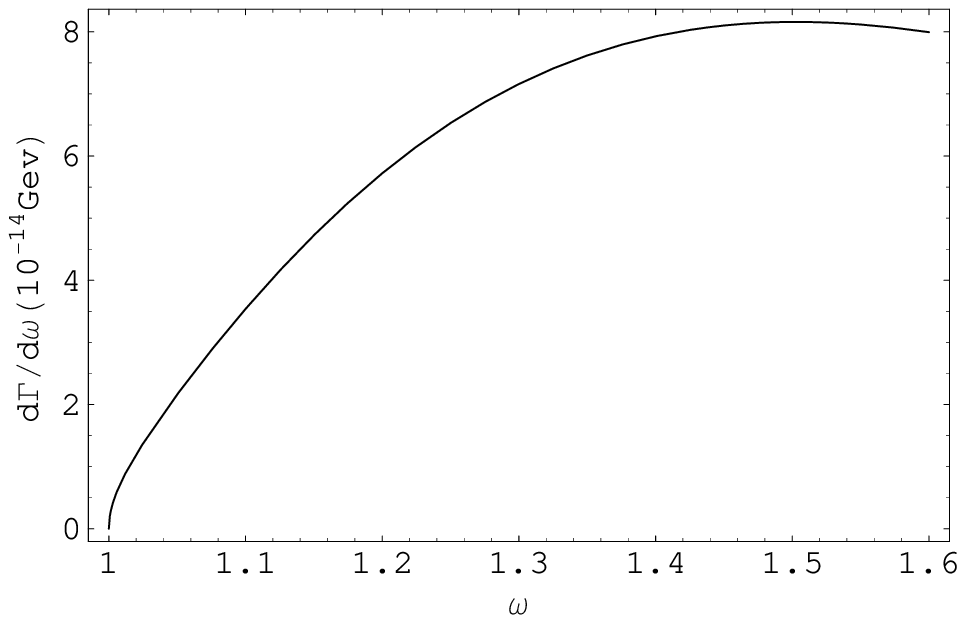}}
\caption {The differential decay rate of $\Omega_b\rightarrow
\Omega_c\ l\ \bar\nu$.}
\end{figure}

\begin{figure}
\scalebox{0.8}[0.8]{\includegraphics{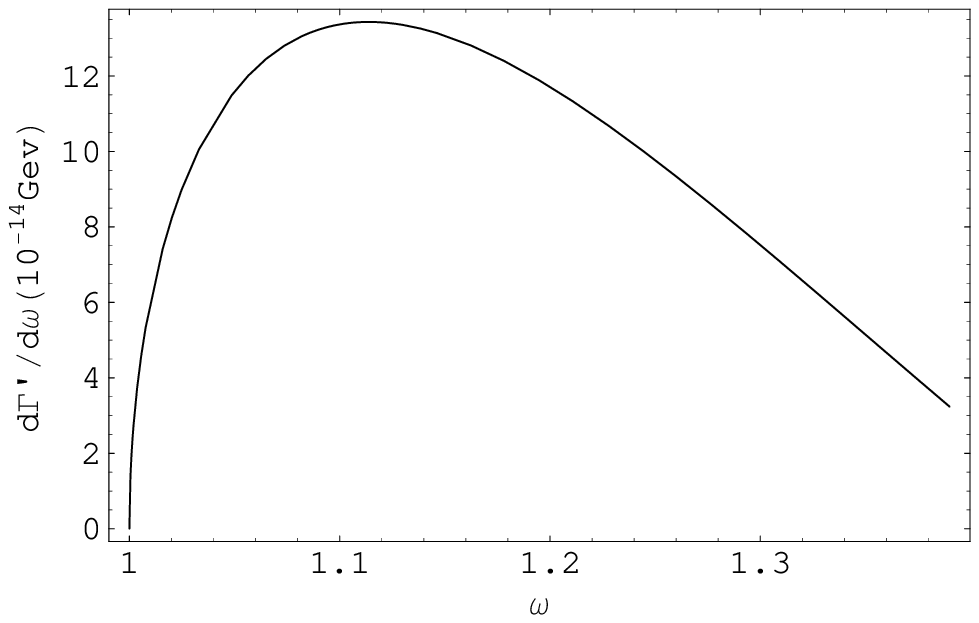}}
\caption {The differential decay rate of $\Omega_b\rightarrow
\Omega^*_c\ l\ \bar\nu$.}
\end{figure}

%%%%%%%%%%%%%%%%%%%%%%%%%%%%%%%%%%%%%%%%%%%%
\section{Summary}
%%%%%%%%%%%%%%%%%%%%%%%%%%%%%%%%%%%%%%%%%%%

In this paper, we have calculated $\Omega_b \rightarrow \Omega_c^{(*)}$
semileptonic decays.  Relevant helicity amplitudes have been written
down.  Both unpolarized and polarized $\Omega_b$ baryon cases have been
considered.  Decay angular distributions, asymmetry parameters and
semileptonic decay rates have been calculated, with numerical results
using leading order results of HQET.  The large $N_c$ QCD result for
Isgur-Wise functions have been used.  The numerical results (especially
the zero-recoil values) can be checked by the experiment at the LHCb.

\acknowledgments

This work was supported in part by the National Natural Science
Foundation of China under nos. 11075193 and 10821504, and by the
National Basic Research Program of China under Grant No. 2010CB833000.

\end{document}